\documentstyle[twoside,fleqn,espcrc2,epsf]{article}

\newcommand{\del}{{\bf D}}

\newcommand{\delv}{{\bf \del}}

\newcommand{\delsq}{D^{(2)}}

\newcommand{\mbare}{\mbox{$M_0$}}

\newcommand{\Ev}{{\bf E}}
\newcommand{\Bv}{{\bf B}}

\newcommand{\sigmav}{\mbox{\boldmath$\sigma$}}
\newcommand{\Sigmav}{\mbox{\boldmath$\Sigma$}}
\newcommand{\gammav}{\mbox{\boldmath$\gamma$}}
\newcommand{\alphav}{\mbox{\boldmath$\alpha$}}

\newcommand{\ainv}{{a^{-1}}}
\newcommand{\be}{\begin{equation}}
\newcommand{\ee}{\end{equation}}

\newcommand{\order}{{\cal O}}

\newcommand{\AmS}{{\protect\the\textfont2
  A\kern-.1667em\lower.5ex\hbox{M}\kern-.125emS}}

\title{ B Meson Decay Constants Using NRQCD }

\author{
        Junko Shigemitsu 
        \address{Physics Department, 
        The Ohio State University, 
        Columbus, Ohio 43210, USA.}
        \thanks{Talk presented at the International Workshop ``Lattice QCD 
on Parallel Computers'', Tsukuba University, March 10 -15, 1997.}
                           }
       
\begin{document}

\begin{abstract}

Recent results for B meson decay constants with NRQCD 
b-quarks and clover light quarks are discussed. 
 Perturbative matching factors through \order($\alpha/M$) are
 now available and incorporated into the analyses. An \order($\alpha \, a$) 
improvement term to the heavy-light axial current is identified and included. 
The slope of $f_{PS}\sqrt{M_{PS}}$ versus $1/M_{PS}$ is significantly 
reduced by these corrections.

\end{abstract}

\maketitle

\section{Introduction}

\noindent
The heavy-light pseudoscalar meson decay constant, $f_{PS}$, is defined 
 through the matrix element of the heavy-light axial vector 
current between the pseudoscalar meson state and the hadronic vacuum.
In Euclidean space one has, 

\be
   \langle \, 0 \,| \, A_{\mu} \,|\,PS\, \rangle = p_{\mu} f_{PS} . 
\ee

\noindent
$f_{D_s}$ has been measured experimentally, giving values consistent with 
lattice predictions.  $f_B$, on the other hand is unlikely to 
be measured directly any time soon via leptonic B decays,
$ B \longrightarrow l \; \overline{\nu}$. 
This fact coupled with the importance 
of $f_B$ in analyses of $B_0 \overline{B_0}$ mixing phenomena,  makes  an 
accurate lattice determination of $f_B$ particularly relevant \cite{flynn96}.

\vspace{.1in}

This talk describes work being carried out by A. Ali Khan, T. Bhattacharya, 
S. Collins, C. Davies, R. Gupta, C. Morningstar, U. Heller, J. Sloan and 
myself,  on heavy meson decay constants with NRQCD b-quarks.  Earlier 
works exist by C. Davies \cite{cdavies93}, Draper \& McNeile \cite{draper}
 and S. Hashimoto \cite{hashimoto}.  Dr. Onogi 
discusses the Hiroshima group's results on NRQCD decay constants 
in a separate talk \cite{onogi}.

\vspace{.1in}
\noindent
New developments since Lattice '96 include,

\begin{itemize}
\item  Completion of the one-loop matching calculation between lattice NRQCD 
and continuum full QCD axial currents through \order($\alpha / M$).  

\item Inclusion of an \order($\alpha \, a$) discretization correction to
 the local heavy-light axial current.  

\item  Considerable decrease in the slope of $f_{PS} \sqrt{M_{PS}}$ 
versus $ 1/M_{PS}$  once  the discretization correction and the matching 
Z-factors have been included.  This makes the connection between the static 
limit and the physical b-quark region much smoother than previously 
thought.

\item  Higher statistics on quenched calculations at $\beta = 6.0$ 
that include $1/M^2$ corrections to the action and currents at 
tree level.  This latter data is still being analyzed and I will 
only  show older data in this talk with corrections through \order($1/M$).

\end{itemize}

\section{ The Action and Current Operators}

\noindent
We use the NRQCD action to simulate b-quarks \cite{nrqcd,cornell}. 
It is given by (in continuum notation),

 \be
\bar{\psi} \left(\del_t  + H_0 + \delta H \right) \psi,
 \ee
We have dropped the rest mass term and 
$H_0$ is the nonrelativistic kinetic energy operator,
 \be
 H_0 = - {\delsq\over2\mbare},
 \ee
and through \order($\Lambda_{QCD} \; (\Lambda_{QCD} / M)$) one has 
\be \label{deltah1}
\delta H  = - \frac{g}{2\mbare}\,\sigmav\cdot\Bv
\ee
In our more recent simulations we have added the terms 
$$
 \frac{ig}{8(\mbare)^2}\left(\delv\cdot\Ev - \Ev\cdot\delv\right) 
- \frac{g}{8(\mbare)^2} \sigmav\cdot(\delv\times\Ev - \Ev\times\delv) $$

\noindent
which are of \order($\Lambda_{QCD} \; (\Lambda_{QCD} / M)^2$) and
\be
- \frac{(\delsq)^2}{8(\mbare)^3}
\ee
which we believe is the dominant \order($\Lambda_{QCD} \; (\Lambda_{QCD}
 / M)^3$) contribution.  These latter calculations 
also include discretization corrections to the lattice laplacian and 
 the lattice time-derivative\cite{cornell}.  All operators in the 
NRQCD action and in the heavy-light currents discussed below, are 
tadpole improved ($ U \rightarrow U/u_0$ with $u_0 =$ fourth root of 
the average plaquette). Tree level coefficients are used in the action.

\vspace{.1in}
\noindent
For light quarks we have data using the Wilson  and the 
$C_{SW} = 1$ clover quark actions.  The bulk of our results,
however,  come from the tadpole improved clover action. Gauge fields 
were created using the Wilson plaquette action. The quenched 
configurations are at $\beta = 6.0$; one set 
 provided by the UKQCD collaboration and another newer set created 
by us on the LANL and NCSA CM5 machines.  The dynamical calculations were 
carried out on  
the HEMCGC $n_f = 2$ staggered configurations at $\beta = 5.6$.  

\vspace{.1in}
\noindent
Heavy-light currents in full QCD have the form  $\bar{q} \Gamma h$.  The 
four component Dirac spinor for the heavy quark, $h$, is related to the 
two component NRQCD heavy quark (heavy anti-quark)
fields, $\psi$ ($\tilde{\psi}$), via an inverse Foldy-Wouthuysen 
transformation.

\be \label{fw}
h = U^{-1}_{FW} \Psi_{FW} = U^{-1}_{FW} \left(\begin{array}{c} 
                                        \psi  \\
                                      \tilde{\psi}
                       \end{array}  \right)  
\ee

\begin{eqnarray}
U^{-1}_{FW} &=& 1 - \frac{1}{2 M}(\gammav \cdot \delv) 
  \nonumber \\
  && + \frac{1}{8 M^2}(\delsq + g \Sigmav \cdot \Bv -2i
 \alphav \cdot \Ev)  \nonumber \\
  && \qquad \qquad \qquad \qquad+ \; \order{(1/M^3)} 
\end{eqnarray}
$\alphav \equiv \gamma_0 \gammav$ and $\Sigmav = 
diag(\sigmav, \sigmav)$.   All our expressions have been converted to
 Euclidean space with hermitian $\gamma$-matrices.\\
Through \order($ 1/M$) one has at tree-level,
\begin{eqnarray} \label{jtree}
J &=& J^{(0)} + J^{(1)}  \nonumber \\
  &=& \bar{q} \Gamma Q - \frac{1}{2 M} \bar{q} \Gamma (\gammav
\cdot \delv) Q
\end{eqnarray}
$Q = \frac{1}{2}(1 + \gamma_0) \Psi_{FW} $.\\
In the next section we discuss what happens at one-loop.

\section{ Perturbative Matching}

\noindent
In order to match between lattice currents used in our simulations and
those of continuum QCD, we consider the process in which a
 heavy quark of momentum $p$ is scattered by the heavylight current into  
a light quark of momentum $ p^{\prime}$.  For a one-loop matching we need
to go through the following steps.

\begin{enumerate}

\item Carry out the one-loop calculation for the above process in full QCD.

\item Expand the amplitude in terms of $ p/M$ , $ p^{\prime}/M$ etc.  

\item Identify operators in the effective theory (NRQCD) that would 
reproduce these $ 1/M$ corrections.  

\item Carry out a one-loop mixing matrix calculation in the effective 
theory.  

\end{enumerate}

\noindent
For HQET analogous calculations have been done by Eichten \& Hill 
\cite{eichtenhill},  
Golden \& Hill \cite{goldenhill},and Neubert \cite{neubert}.
  Colin Morningstar and I have now completed 
the matching calculation, between lattice NRQCD and full continuum 
QCD, for the time component of the axial vector current through 
\order($\alpha/M$) \cite{cmjs}. 
 
\vspace{.1in}
\noindent
\underline{ Continuum  calculation}

\vspace{.1in}
\noindent
A one-loop calculation in full QCD finds that \\
 $\langle \, q(p^{\prime}) \,| \, A_0 \, | \,h(p) \, {\rangle}_{QCD}$ 
involves terms proportional to, 

\begin{eqnarray}
& & \bar{u}_q(p^{\prime}) \gamma_5 \gamma_0 u_h(p) \quad , \quad
\frac{p_0}{M}\;\bar{u}_q(p^{\prime}) \gamma_5  u_h(p) \nonumber  \\
& & \frac{p \cdot p^{\prime}}{M^2}\;
\bar{u}_q(p^{\prime}) \gamma_5 \gamma_0  u_h(p) \quad , \quad
\frac{p_0^{\prime}}{M}\;\bar{u}_q(p^{\prime}) \gamma_5  u_h(p) 
\nonumber \\
& & \frac{p \cdot p^{\prime}}{M^2} \frac{p_0}{M} \;
\bar{u}_q(p^{\prime}) \gamma_5  u_h(p) \quad + \quad \order(1/M^2)
\end{eqnarray}
We set the light quark mass 
equal to zero, i.e. we ignore terms such as $m_q/M$.  \\
The above terms can be reproduced  via three operators in the 
effective theory, $J^{(i)}_A = \bar{q}\, O^{(i)}_A \, Q$ ,\\ 
 i = 0,1,2 , after making use of equations of motion for the light quark. 

\begin{eqnarray} \label{jop}
J^{(0)}_A &=& \bar{q} \gamma_5 \gamma_0 Q  \nonumber \\
J^{(1)}_A &=& - \frac{1}{2M}\,\bar{q} \gamma_5 \gamma_0
(\gammav \cdot \delv) Q  \nonumber \\
J^{(2)}_A &=& \frac{1}{2M} \, (\delv\bar{q} \cdot \gammav ) \gamma_5
 \gamma_0 Q 
\end{eqnarray}
$J^{(0)}$ and $J^{(1)}$ coincide with eq.(\ref{jtree}), 
 $J^{(2)}$ appears only 
at one-loop.  So, through one-loop 
 $\langle \, q(p^{\prime}) \,| \, A_0 \, | \,h(p) \, {\rangle}_{QCD}$ 
can be written as,

\begin{eqnarray} \label{ajren}
\lefteqn{\langle \, q(p^{\prime}) \,| \, A_0 \, | \,h(p) \, 
{\rangle}_{QCD}}  \nonumber \\
&=& \sum^2_{i=0} \eta^A_i \langle \, J_A^{(i)} \,\rangle_{ren} \nonumber \\
&\equiv& \sum^2_{i=0} \eta^A_i \left( \bar{u}_q(p^{\prime}) 
\tilde{O}^{(i)}_A U_Q(p) \right)
\end{eqnarray}
 
\vspace{.1in}
\noindent
$\tilde{O}^{(i)}_A$ are momentum space representations of the operators 
 between quark fields in eq.(\ref{jop}). 
We have calculated the one-loop coefficients, $\eta^A_i$, using dimensional 
regularization (with totally anti-commuting $\gamma_5$) in the 
$\overline{MS}$ scheme. 
A gluon mass, $\lambda$, is introduced as an infrared regulator.
 We find,

\begin{eqnarray} \label{eta}
\eta^A_0  &=& 1 + \frac{\alpha}{3 \pi} \left[ 3\, ln\frac{M}{\lambda}
- \frac{3}{4} \right]  \nonumber \\
\eta^A_1  &=& 1 + \frac{\alpha}{3 \pi} \left[ 3\, ln\frac{M}{\lambda}
- \frac{19}{4} \right]  \nonumber \\
\eta^A_2  &=&  \qquad \frac{\alpha}{3 \pi} \left[12 - \frac{16 \pi}{3}
 \frac{M}{\lambda} \right] 
\end{eqnarray}
The infrared divergent terms will cancel when we match to a lattice 
regularized one-loop calculation.
\vspace{.1in}

\noindent
\underline {Lattice calculation}

\vspace{.1in}
\noindent
The lattice currents used in our simulations are discretized versions  
 of the operators in eq.(\ref{jop}). In the absence of any improvement,
 $J_{A,L}^{(0)}$ becomes a local heavy-light current, and 
$J_{A,L}^{(1)}$ and $J_{A,L}^{(2)}$ have $\delv \rightarrow $ \{lattice 
symmetric covariant derivative\}.  These lattice operators are only defined 
up to improvement terms.  In fact, in the next section we will 
argue that a consistent
one-loop calculation with clover (i.e. \order($a$) improved) light quarks, 
requires that an \order($\alpha \, a$) correction term be added to 
the local $J_{A,L}^{(0)}$.  \\
The lattice matrix elements
 $\langle \, J_{A,L}^{(i)} \, \rangle$ 
can be related to the
 $\langle \, J_A^{(i)} \, \rangle_{ren}$
 on the RHS of eq.(\ref{ajren}) via a mixing matrix ${\bf Z }$.

\be \label{lvren}
 \langle \, J_{A,L}^{(i)} \, \rangle = \sum_j Z_{ij} \, 
 \langle \, J_A^{(j)} \, \rangle_{ren}
\ee
From (\ref{ajren}) and (\ref{lvren}) one extracts the matching relation 
between matrix elements in full QCD and those evaluated in lattice 
simulations,

\begin{eqnarray} \label{ajlat}
{\langle  \, A_0 \, \rangle}_{QCD}
&=& \sum_{i,j} \eta^A_i Z_{ij}^{-1} 
 \langle \, J_{A,L}^{(j)} \, \rangle   \nonumber \\
&\equiv& \sum_j C_j \langle \, J_{A,L}^{(j)} \, \rangle   
\end{eqnarray}
$\vec{C} = \vec{\eta}^A \, {\bf Z}^{-1}$ is the vector of matching
 coefficients for this problem. \\
We have calculated the one-loop contributions to $Z_{ij}$ in lattice 
perturbation theory for lattice actions including $H_0$ and $\delta H$ 
of eq.(\ref{deltah1}).  Equation (\ref{ajlat}) can be expanded out as,

\vspace{.1in}
\noindent
$\langle \, A_0 \,  \rangle_{QCD}  = $

\vspace{.07in}
\noindent
$\;\;\; \left( 1 + \alpha \left[ B_0 - \frac{1}{2}(C_q + C_Q) -
\zeta_{00} - \zeta_{10}\right]\right)\langle J_{A,L}^{(0)}\rangle $

\vspace{.07in}
\noindent
$+ \left( 1 + \alpha \left[ B_1 - \frac{1}{2}(C_q + C_Q) -
\zeta_{01} - \zeta_{11}\right]\right)\langle J_{A,L}^{(1)}\rangle $

\be \label{ajalpha1}
 + \;  \alpha \left[ B_2 -
\zeta_{02} - \zeta_{12}\;\right]\langle J_{A,L}^{(2)}\rangle
 \quad + \quad \order(\alpha^2)
\ee
The $B_i$'s come from the $\eta^A_i$'s in (\ref{eta}) and $C_q$ and $C_Q$ 
are the light and heavy quark lattice wave function renormalizations
 respectively.  The $\zeta_{ij}$ are the one-loop vertex correction 
contributions to $Z_{ij}$.  Before proceeding with combining perturbative 
numbers with simulation results for the matrix elements $\langle \,
 J_{A,L}^{(i)} \, \rangle$, we must discuss improvement of 
the local current $J_{A,L}^{(0)}$.

\section{ An \order($\alpha \, a$) Correction to the Heavy-Light Axial Current}

There has been a lot of work recently on improving 
quark bilinear operators such as the vector and axial currents.  For 
the light quark sector, the DESY \cite{desy} group has shown that an improved 
 axial vector current takes on the form (suppressing isospin 
indices),

\be  \label{aimp}
A^I_{\mu} = A^{loc}_{\mu} + c_A\;a \, D_{\mu} P 
\ee
with $P$ the pseudoscalar density and $D_{\mu}$ the symmetric 
lattice derivative.   In perturbation theory, 
$c_A$ starts out at \order($\alpha$) \cite{desy,heatlie}. 
 Less is known about improvement of 
the local heavy-light or static-light axial currents.  Borrelli \& Pittori 
\cite{borrelli} have shown that through one-loop, static-light bilinears 
 using clover light quarks are free of 
\order($a$) and \order($\alpha \; a\,log(a)$)  terms.  They did not consider 
\order($\alpha \; a$) terms.  \\
Our one-loop calculation of the mixing matrix $Z_{ij}$  shows that
 there is an 
\order($\alpha \; a$)  lattice artifact term whose effects can be 
removed by improving $J_{A,L}^{(0)}$.  The improved heavylight 
lattice axial current acquires a correction term that is the precise 
analogue of the correction term in eq.(\ref{aimp}).
The argument goes as follows.

\vspace{.1in}
\noindent
In order to calculate $Z_{02} = \alpha \, \zeta_{02}$ one needs to
 start from $\langle \,
 J_{A,L}^{(0)} \, \rangle$ and project out terms proportional to 
$ \langle \, J^{(2)}_A \, \rangle_{ren} =
\frac{1}{2M} \, \bar{u}_q (-i {\bf p}^{\prime} \cdot \gammav ) \gamma_5
 \gamma_0 \, U_Q $.   \\
After calculating $\zeta_{02}$ in this way, we find that $\zeta_{02}$ 
has a term that grows with $a \, M$, the dimensionless heavy quark mass. 
So, 

$\alpha \, \zeta_{02} \, \langle \, J_A^{(2)} \, \rangle_{ren} =  
\alpha \, ( aM\, A_1 + A_2) \, \langle \, J_A^{(2)} \, \rangle_{ren}  $

\noindent
has an \order($\alpha \, a$) term with the structure,

$\alpha \, a \, 
 \bar{u}_q (-i {\bf p}^{\prime} \cdot \gammav ) \gamma_5
 \gamma_0 \, U_Q $.  

\noindent
One can interpret this term as coming from a lattice operator, 

\be \label{aj3}
J^{(3)}_{A,L} = a \, (\delv\bar{q} \cdot \gammav ) \gamma_5 \gamma_0 Q 
\ee
and what we are finding is,

\begin{eqnarray}
\lefteqn{
\alpha \, \zeta_{02} \, \langle \, J_A^{(2)} \, \rangle_{ren}}
\nonumber \\
&=& \alpha \, \zeta_{02}^{true} \, \langle \, J_A^{(2)} \, 
\rangle_{ren} + \alpha \, \zeta_{03} \, \langle \, J_{A,L}^{(3)} \, 
\rangle_{ren}
\end{eqnarray}
and from eq.(\ref{lvren}),
\begin{eqnarray} \label{lvren0}
 \langle \, J_{A,L}^{(0)} \, \rangle &= &\sum_{j=0,1,2}Z_{0j} \, 
 \langle \, J_A^{(j)} \, \rangle_{ren} \nonumber \\
&&\quad \quad \quad + \; \alpha \; \zeta_{03} \, \langle \, J_{A,L}^{(3)} \, 
\rangle_{ren} 
\end{eqnarray}
with $Z_{02} = \alpha \, \zeta_{02}^{true}$.  
If one wants to have the same set of matrix elements 
 on the RHS of eq.(\ref{lvren0})
as in the continuum theory, the last term must be removed.  This is 
easily accomplished by improving $J_{A,L}^{(0)}$.

\be
J_{A,L}^{(0)} \rightarrow J_{A,L}^{(0),I}
 = J_{A,L}^{(0)} - \alpha \, \zeta_{03} J_{A,L}^{(3)}
\ee
and one now has, 
\be 
 \langle \, J_{A,L}^{(0),I} \, \rangle = \sum_{j=0,1,2}Z_{0j} \, 
 \langle \, J_A^{(j)} \, \rangle_{ren}
\ee
Going through the same steps that led from (\ref{lvren}) to (\ref{ajlat}) 
and then to (\ref{ajalpha1}), one ends up with an equation very similar 
to eq.(\ref{ajalpha1}) with \\
$\langle\,J_{A,L}^{(0)}\,\rangle \rightarrow \langle \,J_{A,L}^{(0),I}
\,\rangle$ and $\; \zeta_{02} \rightarrow \zeta_{02}^{true}$.  

\vspace{.1in}
\noindent
Although $J_{A,L}^{(2)}$ and $J_{A,L}^{(3)}$ are proportional to 
each other, 
 $J_{A,L}^{(3)} = 2 \, aM \,J_{A,L}^{(2)}$,
they play very different roles.
 $J_{A,L}^{(2)}$ is the lattice version of a current operator that exists in 
the continuum theory.  It is a $1/M$ correction to the static heavy-light axial
current and is absent in the static theory.  
 $J_{A,L}^{(3)} $, on the other hand, has no continuum counter part.  It 
survives into the lattice static theory.  So the static limit of 
eq.(\ref{ajalpha1}) becomes,

\vspace{.1in}
\noindent
$\langle \, A_0 \,  \rangle_{QCD}  $

\vspace{.07in}
\noindent
$= \left( 1 + \alpha \left[ B_0 - \frac{1}{2}(C_q + C_Q) -
\zeta_{00} \right]\right)\langle J_{A,L}^{(0),I}\rangle_{stat} $

\vspace{.08in}
\noindent
$= \left( 1 + \alpha \left[ B_0 - \frac{1}{2}(C_q + C_Q) -
\zeta_{00} \right]\right)\langle J_{A,L}^{(0)}\rangle_{stat} $

\be \label{ajalphst}
\quad - \;  \alpha \,\;
\zeta_{03}\langle J_{A,L}^{(3)}\rangle_{stat}
 \quad + \quad \order(\alpha^2)
\ee
We will see in the next section that the last correction term in 
(\ref{ajalphst}) significantly reduces the value of 
$f_{PS} \sqrt{M_{PS}}$ in the static theory.  

\vspace{.1in}
\noindent
It is easy to see that the improvement term $J_{A,L}^{(3)}$ 
 is of the same form as the second term in eq.(\ref{aimp}).  If one defines 
a heavylight pseudoscalar density $P_{HL} \equiv \bar{q} \gamma_5 Q$, then 

\begin{eqnarray} \label{d0aj3}
\alpha \, a \, D_0 P_{HL} &=& \alpha \, a \, (D_0\bar{q}\,) \gamma_5 Q \; + 
 \; \order( \alpha\, a/M)  \nonumber \\
&\approx& - \alpha \, a \, (D_0 \bar{q}\, \gamma_0) \gamma_5 \gamma_0 Q
\end{eqnarray}
In the above we ignore the term coming from the time-derivative acting 
on the heavy quark field $Q$, since equations of motion make that 
into an \order($\alpha\, a/M$) term and we neglect such contributions 
together with contributions of \order($\alpha/M^2$).  Applying light quark 
equations of motion to the last expression in (\ref{d0aj3}) gives the 
operator $J_{A,L}^{(3)}$ of eq.(\ref{aj3}) multiplied by $\alpha$.

\section{ Some Quenched $f_B$ Results}
An analysis of NRQCD heavy meson decay constants including one-loop 
matching factors,  has been carried out for the first time in 
Ref.\cite{arifa}. 
The data was obtained on $16^3 \times 48$ quenched configurations at 
$\beta = 6.0$.  Both the gauge configurations and light propagators were 
generously  provided by the UKQCD collaboration.  Tadpole improved clover 
light propagators  were used.  The NRQCD
action included the $\delta H$ of eq.(\ref{deltah1}) but no $1/M^2$ terms.  
  Two light $\kappa$ values around the strange quark mass and 
four NRQCD bare heavy quark masses were used.  We also have results for 
static heavy quarks.  

\vspace{.1in}
\noindent
For the perturbation theory, we use $\alpha_V$ of 
Lepage\&Mackenzie\cite{lepmac}.  The 
$q^*$ for this matching calculation has not been calculated yet.  So, we 
present results for both $q^* = 1/a$ and $q^* = \pi/a$.  The $ log(aM)$ terms 
that appear in the matching coefficients after cancellation of logarithmic 
IR divergences, are set to a constant value $ log(a \overline{m}_b)$ for 
all bare heavy quark masses.  $ \overline{m}_b = 4.1(1) GeV$ is 
the b-quark $\overline{MS}$ mass at a scale equal to its value
 \cite{mb}.  Alternatively we could have used the 
one-loop renormalization group improved expression with 
an overall factor of $\left(\alpha(\overline{m}_b)/\alpha(\ainv)\right)^{-2/
\beta_0}$ for the two matching coefficients $C_0$ and $C_1$.  The difference 
between the two approaches is very slight and at the B-meson 
numerically undetectable.  Since we presently
  have results at only one lattice 
spacing, we have used the simpler $log (a \overline{m}_b)$ prescription.
  Details of the simulations and fitting procedures are given 
in Ref.\cite{arifa}.
  Figure 1. shows $a^{3/2}f_{PS} \sqrt{M_{PS}}$
 versus $ 1/aM_{PS}$  for $\kappa = \kappa_s$. 
 We show the tree-level results and the one-loop 
results for the two $q^*$'s.  For the static limit we also indicate 
the one-loop values without the $\langle\,J_{A,L}^{(3)}\,\rangle$ 
correction term.  The physical $B_s$ meson is just below 
$1/aM_{PS} = 0.4$ on the figure.

\begin{figure}[t]
\epsfxsize=9.0cm
\centerline{\epsfbox{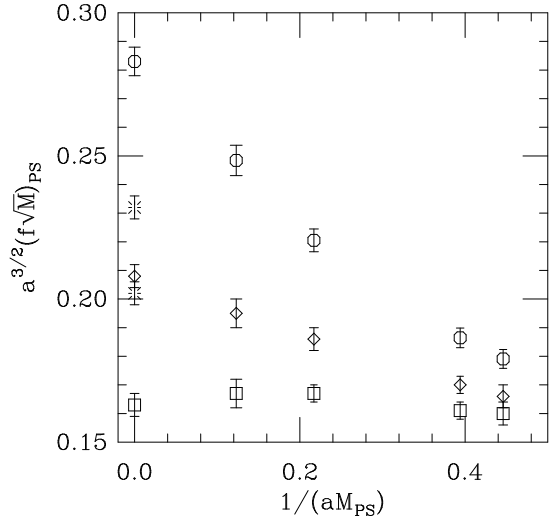}}
\caption{  Heavy-Light Decay Constants 
at $\kappa_s$.
 Circles : tree-level ; Diamonds : one-loop matching with $aq^* = \pi$ ;
Squares : one-loop matching with $aq^* = 1$.  We demonstrate the effect of 
the correction term to $J_{A,L}^{(0)}$, by showing the one-loop matched 
result in the static limit without the last term in eq.(\ref{ajalphst})
 for $aq^* = \pi$ (upper burst) 
and $aq^* = 1$ (lower burst).
 The physical $B_s$ meson 
is just below $1/aM_{PS} = 0.4$. }

\end{figure}

\vspace{.1in}
\noindent
There are several interesting things to note in Figure 1. 

\vspace{.1in}
\noindent
1. After including the one-loop matching factors and improvement of 
$J_{A,L}^{(0)}$,  the slope of $a^{3/2}f_{PS} \sqrt{M_{PS}}$ versus
 $ 1/aM_{PS}$ 
as one leaves the static limit decreases considerably.  For $q^* = 1/a$ 
it is consistent with zero.  For $q^* = \pi/a$ the relative slope 
( the slope divided by $\left( f_{PS}\sqrt{M_{PS}} \right)_{stat}$ ) 
is $ \sim -1 GeV$.  Unfortunately a precise value of 
$q^*$ is required in order to make 
a more quantitative statement.  Ref.\cite{hernan} quotes $q^* = 2.18/a$ for 
the static theory.  

\vspace{.1in}
\noindent
2. Around the physical $B$ meson region the one-loop corrections are a
9 - 13\%  effect depending on $q^*$ at $\kappa_s$.  The difference 
due to $q^*$ of $1/a$ or $ \pi/a$ is a $\sim$5\% effect on the final answer.
This is one measure of the uncertainty in $f_{B_s}$ coming from 
higher orders in the  matching calculation. \\
 The static limit is 
much more sensitive to $q^*$ and the one-loop corrections are large
 (25 - 40\% at $\kappa_s$ and 10 - 30\% at $\kappa_c$), with a third of 
the shift coming from the \order($\alpha \, a$) correction to $J_{A,L}^{(0)}$.

\vspace{.1in}
\noindent
We use a scale of $\ainv$ = 2.0(2)GeV to convert to physical units.  In a 
quenched calculation, different observables can lead to different 
$\ainv$'s.  We have used an $\ainv$ consistent with quenched light quark 
calculations ($M_{\rho}$, $f_{\pi}$ etc.) and allow for a large 
error of 10\%.  This feeds back into one of the dominant systematic 
errors in our final estimate for $f_B$.  We find at $\kappa_s$,

\[f_{B_s} =
\cases{ 
0.198\,(8)(30)(17)\; GeV  \qquad q^* = 1/a  \cr
0.209\, (8)(32)(17)\; GeV  \qquad q^* = \pi/a \cr} \]

\noindent
and after extrapolating to $\kappa_c$,

\[f_{B} =
\cases{ 
0.174\,(28)(26)(16)\; GeV  \qquad q^* = 1/a  \cr
0.183\, (32)(28)(16)\; GeV  \qquad q^* = \pi/a \cr} \]

\noindent
Errors in the first brackets correspond to statistical plus fitting plus 
$\kappa$ extrapolation errors.  In the second brackets we give $\ainv$ 
systematic errors and the third brackets summarize our estimates for 
higher order perturbative and $1/M^2$ corrections.  We now have more 
recent simulations with $1/M^2$ terms included at tree level both in the 
NRQCD action and the currents (see next section on Work in Progress).
  Their contributions are at the 3 - 4\% level.
 The Hiroshima group reports similar findings 
for the $1/M^2$ current corrections \cite{onogi}. 
We have not included any estimate for continuum extrapolation 
corrections, since at the moment we only have results at a single value
of $\beta$.  Studies with Wilson and/or static fermions 
\cite{eichten,milc,jlqcd} have shown 
noticeable lattice spacing dependence around $\beta = 6.0$.  It 
will be interesting to do a thorough scaling study with clover light fermions 
and an improved $J_{A,L}^{(0)}$.

\section{Work in Progress and Future Plans}

Two sets of NRQCD heavy meson decay constant 
data are currently being analyzed by my collaborators.  

\vspace{.1in}
\noindent
 1. The Glasgow-LANL-OSU-Kentucky (GLOK) 
collaboration has results on $16^3 \times 48$ quenched configurations at 
$\beta = 6.0$ \cite{arifa2}.
  Both the NRQCD action and the current operators include 
$1/M^2$ corrections at tree level.  We also use a better time 
evolution equation to obtain heavy propagators ( which follows from a 
slightly modified lattice NRQCD action) than in the simulations 
of the previous section.  The new evolution equation eliminates 
a residual \order($ a\Lambda \, (\Lambda/M)$) error in amplitudes. 
The light quark action employed is the tadpole-improved clover action. 
We use five $\kappa$ values and six heavy quark masses.  Perturbative 
matching calculations for the new lattice action have now also been 
completed \cite{cmjs}.  Some preliminary tree-level results were 
presented at St. Louis \cite{alikhan96}.  

\vspace{.1in}
\noindent
2. The SCRI-Glasgow-OSU (SGO) collaboration has results on 
 the HEMCGC $n_f = 2$, $am_{dyn} = 0.01$ dynamical staggered configurations 
at $\beta = 5.6$.  
Both Wilson and tadpole-improved clover light quarks have been combined with 
NRQCD heavy quarks.  The NRQCD action and the heavy-light currents were 
corrected through \order($1/M$).
  Three $\kappa$ values and eleven heavy quark mass 
values were used.
  The Wilson light fermion results are published in Ref.\cite{collins1}.
Tree level analyses of the clover light quark data have been reported on
in \cite{sara96}.  We are now completing the analysis including 
one-loop matching \cite{collins2}.

\vspace{.1in}
\noindent
Several matching coefficient projects are also on the agenda.  The one-loop
matching calculation for the heavy-light vector current is well 
underway and   $q^*$'s for both axial and vector currents are high on our 
list.
We have also started studies of nonperturbative renormalization of 
NRQCD operators. \\
Finally, quenched simulations at $\beta = 5.7$ have begun 
and in the future we plan to go onto $\beta = 6.2$.

\section{Summary}

The NRQCD approach to B meson decays is looking very promising.  
The $1/M$ expansion is working well and appears to be under good 
control.  The first 
one-loop perturbative matching calculations have been completed and 
incorporated into our analyses.  Uncertainties due to higher orders 
in perturbative matching are estimated to be at the $\sim$5\% level 
around the B meson. 
Many more results should be forthcoming soon, including the use of an
improved heavy quark time evolution equation, and studies of $1/M^2$ 
corrections, unquenching and scaling.  

\vspace{.2in}
\noindent
\underline{Acknowledgements}

I thank the organizers for a very stimulating workshop and all my 
collaborators on the NRQCD heavy-light projects for their support and 
insights. 
Special thanks are due to Arifa Ali Khan for creating and analyzing 
 the NRQCD data presented here plus many discussions on the 
interpretation of the one-loop matched results,  and to Colin Morningstar 
for an enjoyable collaboration on the perturbative calculations.
I have also benefit from conversations with Peter Lepage.  
This work is supported in part by a grant from the US Department of 
Energy DE-FG02-91ER40690 and by NATO CRG941259.


\begin{thebibliography}{99}

\bibitem {flynn96}
For recent reviews see J.D.Richman, talk presented at the 28th ICHEP,
 July 1996, Warsaw, hep-ex/9701014; 
J.Flynn, Nucl. Phys. B (Proc. Suppl.){\bf 53},
 168 (1997).

\bibitem {cdavies93}
UKQCD Collaboration, C.T.H.Davies, Nucl. Phys. B (Proc. Suppl.){\bf 30},
 437 (1994).

\bibitem {draper}
T.Draper and C.McNeile, Nucl. Phys. B (Proc. Suppl.){\bf 47}, 429 (1996).

\bibitem {hashimoto}
S.Hashimoto, Phys. Rev. D {\bf 50}, 4639 (1994).

\bibitem {onogi}
T.Onogi, talk presented at this conference.

\bibitem {nrqcd}
G.P.Lepage and B.Thacker, Nucl. Phys. B (Proc. Suppl.){\bf 4}, 199 (1988).
\bibitem {cornell}
G.P.Lepage et al.,
Phys.\ Rev.\ D {\bf 46}, 4052 (1992).

\bibitem {eichtenhill}
E.Eichten and B.Hill, Phys. Lett. B{\bf 240}, 193 (1990).

\bibitem {goldenhill}
M.Golden and B.Hill, Phys. Lett. B{\bf 254}, 225 (1991).

\bibitem {neubert}
M.Neubert, Phys. Rev. D {\bf 49}, 1542 (1994).

\bibitem {cmjs}
C.Morningstar and J.Shigemitsu, in preparation.

\bibitem {desy}
M.L\"{u}scher, S.Sint, R.Sommer and P.Weisz, Nucl. Phys. B{\bf 478},
 365 (1996).

\bibitem {heatlie}
G.Heatlie et al., Nucl. Phys. B{\bf 352}, 266 (1991).

\bibitem {borrelli}
A.Borrelli and C.Pittori, Nucl. Phys. B{\bf 385}, 502 (1992).

\bibitem {arifa}
A.Ali Khan et al., ``Heavy-Light Mesons with Quenched Lattice NRQCD : 
Results on Decay Constants'', submitted to Phys. Rev. D, hep-lat/9704008.


\bibitem {lepmac}
 G.P.Lepage, P.B.Mackenzie, Phys. Rev. D {\bf 48}, 2250 (1993).

\bibitem {mb}
C.T.H.Davies et al., Phys. Rev. Lett. {\bf 73}, 2654 (1994).

\bibitem {hernan}
O.Hernandez and B.Hill, Phys. Rev. D {\bf 50}, 495 (1994).

\bibitem {eichten}
A.Duncan et al., Phys. Rev. D {\bf 51}, 5101 (1995).

\bibitem {milc}
MILC Collaboration, C.Bernard et al., Nucl. Phys. B (Proc. Suppl.){\bf 53},
 358 (1997); S.Gottlieb, talk presented at this conference.


\bibitem{jlqcd}
JLQCD Collaboration, S.Aoki et al., Nucl. Phys. B (Proc. Suppl.){\bf 53}, 
355 (1997);  S. Hashimoto, talk presented at this conference.

\bibitem {arifa2}
A.Ali Khan et al., in preparation.

\bibitem {alikhan96}
A.Ali Khan and T.Bhattacharya, Nucl. Phys. B (Proc. Suppl.){\bf 53},
 368 (1997).

\bibitem {collins1}
S.Collins et al., Phys. Rev. D {\bf 55}, 1630 (1997).

\bibitem {sara96}
S.Collins, Nucl. Phys. B (Proc. Suppl.){\bf 53}, 389 (1997).

\bibitem {collins2}
S.Collins et al., in preparation.   


\end{thebibliography}
\end{document}